\documentclass[RNAAS]{aastex63}
 	\usepackage{newtxtext,newtxmath}

\newcommand{\Tycho}{Tycho2}

\newcommand{\niceurl}[1]{\href{#1}{#1}}

\begin{document}

\title{Preliminary Target Selection for the DESI Bright Galaxy Survey (BGS)}


\author[0000-0001-8848-9768]{Omar Ruiz-Macias}
\affiliation{Institute for Computational Cosmology, Department of Physics, Durham University, South Road, Durham DH1 3LE, UK}
\affiliation{Institute for Data Science, Durham University, South Road, Durham DH1 3LE, UK}

\author[0000-0002-7305-9578]{Pauline Zarrouk}
\affiliation{Institute for Computational Cosmology, Department of Physics, Durham University, South Road, Durham DH1 3LE, UK}

\author[0000-0002-5954-7903]{Shaun Cole}
\affiliation{Institute for Computational Cosmology, Department of Physics, Durham University, South Road, Durham DH1 3LE, UK}

\author[0000-0002-5875-0440]{Peder Norberg}
\affiliation{Institute for Computational Cosmology, Department of Physics, Durham University, South Road, Durham DH1 3LE, UK}
\affiliation{Centre for Extragalactic Astronomy, Department of Physics, Durham University, South Road, Durham DH1 3LE, UK}

\author[0000-0002-9935-9755]{Carlton Baugh}
\affiliation{Institute for Computational Cosmology, Department of Physics, Durham University, South Road, Durham DH1 3LE, UK}
\affiliation{Institute for Data Science, Durham University, South Road, Durham DH1 3LE, UK}

\author[0000-0002-8458-5047]{David Brooks}
\affiliation{Department of Physics and Astromomy, University College London, Gower Street, London WC1e 6BT} 

\author[0000-0002-4928-4003]{Arjun Dey}
\affiliation{NSF’s National Optical-Infrared Astronomy Research Laboratory, 950 N. Cherry Ave., Tucson, AZ 85719, USA}

\author[0000-0002-2611-0895]{Yutong Duan}
\affiliation{Physics Department, Boston University, Boston, MA 02215, MA}

\author[0000-0002-8281-8388]{Sarah Eftekharzadeh}
\affiliation{Department of Physics and Astronomy, The University of Utah, 115 South 1400 East, Salt Lake City, UT 84112, USA}

\author{Daniel J. Eisenstein}
\affiliation{Harvard-Smithsonian Center for Astrophysics, 60 Garden Street, Cambridge, MA 02138, USA}

\author[0000-0002-2890-3725]{Jaime E. Forero-Romero}
\affiliation{Departamento de Física, Universidad de los Andes, Cra. 1 No. 18A-10, CP 111711, Bogotá, Colombia}

\author[0000-0001-9632-0815]{Enrique Gazta\~naga}
\affiliation{Institute of Space Sciences (ICE, CSIC), 08193 Barcelona, Spain}
\affiliation{
Institut d\'~Estudis Espacials de Catalunya (IEEC), 08034 Barcelona, Spain
}

\author{ChangHoon Hahn}
\affiliation{Lawrence Berkeley National Laboratory, 1 Cyclotron Road, Berkeley, CA 94720, USA}
\affiliation{Berkeley Center for Cosmological Physics, UC Berkeley, CA 94720, USA}

\author{Robert Kehoe}
\affiliation{Department of Physics, Southern Methodist University, 3215 Daniel Avenue, Dallas, TX 75275, USA}

\author[0000-0003-1838-8528]{Martin Landriau}
\affiliation{Lawrence Berkeley National Laboratory, 1 Cyclotron Road, Berkeley, CA 94720, USA}

\author{Dustin Lang}
\affiliation{Perimeter Institute for Theoretical Physics, 31 Caroline Street N, Waterloo, Ontario, N2L 2Y5, Canada}
\affiliation{Department of Physics and Astronomy, University of Waterloo, Waterloo, ON N2L 3G1, Canada}

\author[0000-0003-1887-1018]{Michael E. Levi}
\affiliation{Lawrence Berkeley National Laboratory, 1 Cyclotron Road, Berkeley, CA 94720, USA}

\author{John Lucey}
\affiliation{Centre for Extragalactic Astronomy, Department of Physics, Durham University, South Road, Durham DH1 3LE, UK}

\author[0000-0002-1125-7384]{Aaron M. Meisner}
\affiliation{NSF’s National Optical-Infrared Astronomy Research Laboratory, 950 N. Cherry Ave., Tucson, AZ 85719, USA}

\author{John Moustakas}
\affiliation{Department of Physics and Astronomy, Siena College, 515 Loudon Road, Loudonville, NY 12211}

\author{Adam D. Myers}
\affiliation{University of Wyoming, 1000 E. University Ave., Laramie, WY 82071, USA}

\author{Nathalie Palanque-Delabrouille}
\affiliation{IRFU, CEA, Universit\'e Paris-Saclay, F-91191 Gif-sur-Yvette, France}

\author{Claire Poppett}
\affiliation{Space Sciences Laboratory at University of California, 7 Gauss Way, Berkeley, CA 94720}

\author{Francisco Prada}
\affiliation{Instituto de Astrof\'isica de Andaluc\'ia (CSIC), Glorieta de la Astronom\'ia, E-18080 Granada, Spain}

\author[0000-0001-5999-7923]{Anand Raichoor}
\affiliation{Institute of Physics, Laboratory of Astrophysics, Ecole Polytechnique F\'{e}d\'{e}rale de Lausanne (EPFL), Observatoire de Sauverny, 1290 Versoix, Switzerland}

\author[0000-0002-5042-5088]{David J. Schlegel}
\affiliation{Lawrence Berkeley National Laboratory, 1 Cyclotron Road, Berkeley, CA 94720, USA}

\author{Michael Schubnell}
\affiliation{Department of Physics, University of Michigan, 450 Church St., Ann Arbor, MI 48109, USA}

\author{Gregory Tarl\'e}
\affiliation{Department of Physics, University of Michigan, Ann Arbor, MI 48109, USA}

\author{David H. Weinberg}
\affiliation{Department of Astronomy and the Center for Cosmology and Astroparticle Physics, The Ohio State University, 140 West 18th Avenue, Columbus OH 43210, USA}

\author{M. J. Wilson}
\affiliation{Lawrence Berkeley National Laboratory, 1 Cyclotron Road, Berkeley, CA 94720, USA}
\affiliation{Berkeley Center for Cosmological Physics, UC Berkeley, CA 94720, USA}

\author{Christophe Y\`eche}
\affiliation{IRFU, CEA, Universit\'e Paris-Saclay, F-91191 Gif-sur-Yvette, France}


\begin{abstract}

The Dark Energy Spectroscopic Instrument (DESI) will execute a nearly magnitude-limited survey of low redshift galaxies ($0.05 \leq z \leq 0.4$, median $z \approx 0.2$).
Clustering analyses of this Bright Galaxy Survey (BGS) will yield the most precise measurements to date of baryon acoustic oscillations and redshift-space distortions at low
redshift.
DESI BGS will comprise two target classes: (i) BRIGHT ($r<19.5$~mag), and (ii) FAINT ($19.5<r<20$~mag).
Here we present a summary of the star-galaxy separation, and different photometric and geometrical masks, used in BGS to reduce the number of spurious targets. The selection results in a total density of $\sim 800$ objects/deg$^2$ for the BRIGHT and $\sim 600$ objects/deg$^2$ for the FAINT selections.
A full characterization of the BGS selection can be found in \cite{2020arXiv200714950R}.
\end{abstract}


\section{Introduction} 
The DESI BGS~\citep{DESI2016} will be a flux-limited $r$-band selected sample of 10 million galaxies to $z=0.4$. The DESI target selection will be done on the Legacy Surveys (LS) imaging \citep{Dey2019}. Results presented here are based on the DR8 release\footnote{\url{http://legacysurvey.org/dr8/}}.
The DESI BGS is expected to have a target density of about $800$ galaxies/deg$^{2}$ in a primary selection with a magnitude limit in the $r$-band of $19.5$ and $600$ galaxies/deg$^{2}$  in an additional sample defined by the magnitude range $19.5 < r < 20$ \citep{Smith:2017tzz}.
Henceforth, we will refer to these BGS samples as BRIGHT and FAINT respectively. 




\section{BGS target selection}


The BGS targets are selected using the three optical fluxes in the Legacy Surveys \citep{2020arXiv200714950R}:

\begin{eqnarray}
& & r < 19.5 \text{ for BRIGHT }, \hspace{1cm} 19.5 < r < 20 \text{ for FAINT } \label{eq:main} \\
& & (-1 < g-r < 4) \label{eq:color_gr} \\
& & (-1 < r-z < 4) \label{eq:color_rz} \\
& & {\rm rfibmag}<
\begin{cases}
22.9 +  (r-17.8)  & \text{for } r < 17.8 \\
22.9 & \text{for } 17.8 < r < 20 
\end{cases} \label{eq:fmc}
\end{eqnarray}

%



where $g$, $r$, and $z$ indicate the extinction-corrected AB magnitudes in the corresponding band (using LS extinction corrections). In addition, we require all targets to be covered by at least one image in each optical band. We remove sources near bright stars, large galaxies, or globular clusters by requiring that LS MASKBITS $1$, $12$ and $13$ are not set as defined in the LS data model. 
The bright star mask (defined by  \texttt{bit}=1) combines stars from {\it Gaia} DR2 \citep{2018A&A...616A...1G} and the \Tycho\ \citep{2000A&A...355L..27H} catalog, corrected for epoch and proper motions. This mask consists of a circular exclusion region with a radius that depends on the magnitude of the star, $m$. The magnitude is either the \Tycho\ \textsc{mag}\_\textsc{vt} or {\it Gaia} $G$-mag with {\it Gaia} $G$-mag taking precedence. Stars fainter than $m=13$ are not masked. The large galaxies mask (\texttt{bit}=12) was defined by the Siena Galaxy Atlas -- 2020 (Moustakas, J.~in prep), an angular diameter-limited sample of galaxies with mostly HyperLeda objects\citep{2014A&A...570A..13M}. To mask around large galaxies and elliptical mask is used defined by the diameter at the $25$ mag/arcsec$^2$ (optical) surface brightness, the ratio of the semi-minor axis to the to semi-major axis and the position angle.
 The globular cluster (GC) mask (\texttt{bit}=13) 
consists of a circular exclusion zone around known GCs from the OpenNGC catalog\footnote{https://github.com/mattiaverga/OpenNGC}.

BGS star-galaxy separation is based on {\it Gaia} DR2.  {\it Gaia} is highly complete for BGS and
has a better point spread function (PSF) 
than ground-based surveys. A galaxy in BGS is defined by  $(G-rr > 0.6) \, \texttt{or} \, (G=0)$ where $G$ is {\it Gaia} $G$-mag and $rr$ is LS $r$-band magnitude (without any extinction correction). We apply 
Equation~\ref{eq:color_gr} and \ref{eq:color_rz} to remove sources that are beyond 
the color range of BGS galaxies. To remove spurious large and low-surface-brightness galaxies, we implement a cut that compares the $r$-band magnitude ($r$) to the predicted $r$-band {\it fiber} magnitude ({\rm rfibmag}), as in Equation~\ref{eq:fmc}. Finally, we impose a minimum quality for the data reduced by \textit{The Tractor}\footnote{https://github.com/dstndstn/tractor} \citep{2016ascl.soft04008L}, 
in $\texttt{FRACMASKED}\_i < 0.4$, $\texttt{FRACIN}\_i  > 0.3$, $\texttt{FRACFLUX}\_i  < 5$, where $i \equiv\textrm{$g$, $r$ or $z$}$. \texttt{FRACIN} is used to select sources for which a large fraction of the model flux lies within the contiguous pixels
to which the model was fitted, \texttt{FRACFLUX} is used to reject objects 
that are swamped by flux from adjacent sources, and \texttt{FRACMASKED} is used to veto objects with a high fraction of masked pixels.

\begin{figure}[ht]
\begin{center}
\includegraphics[scale=0.38]{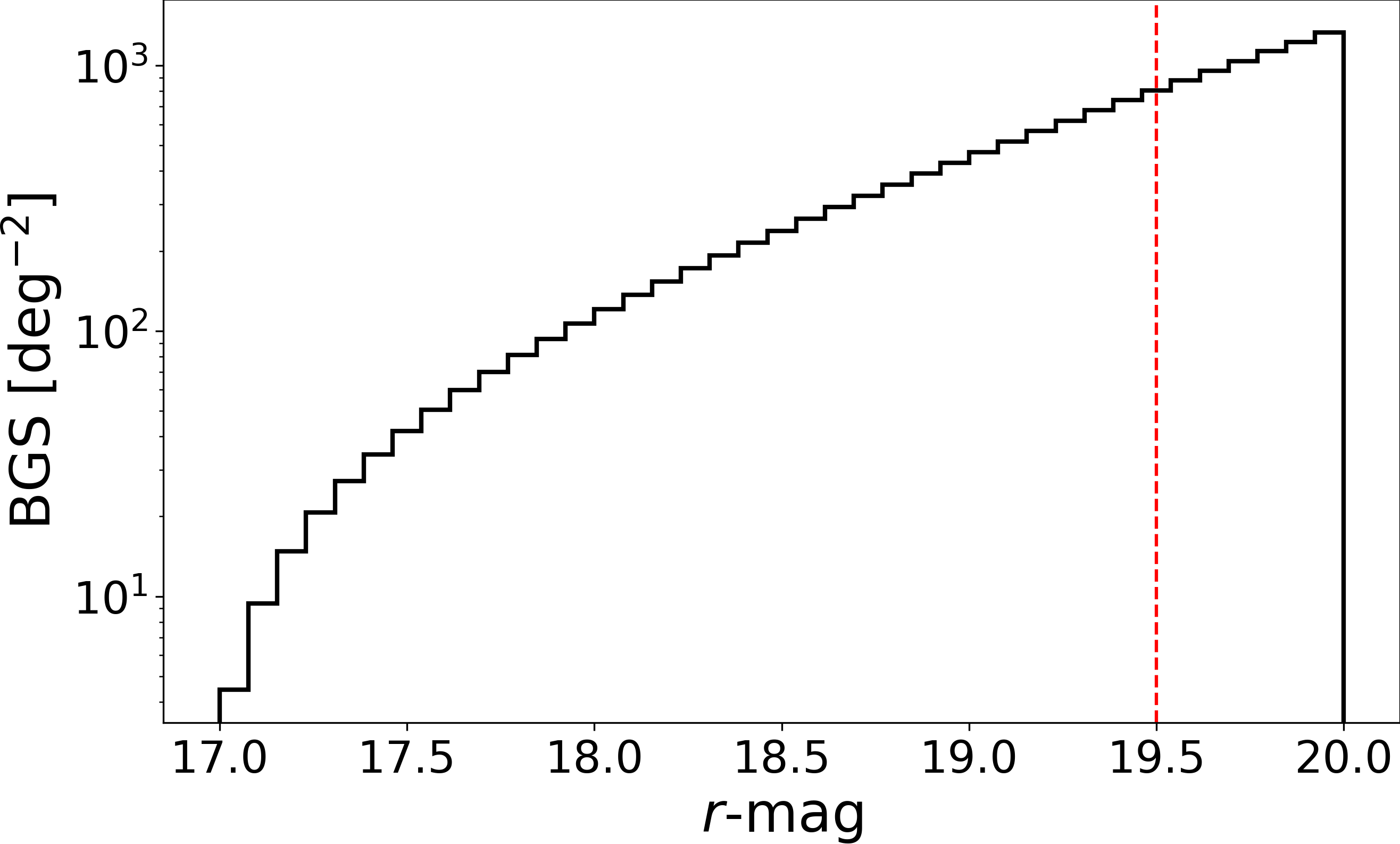}  \includegraphics[scale=0.38]{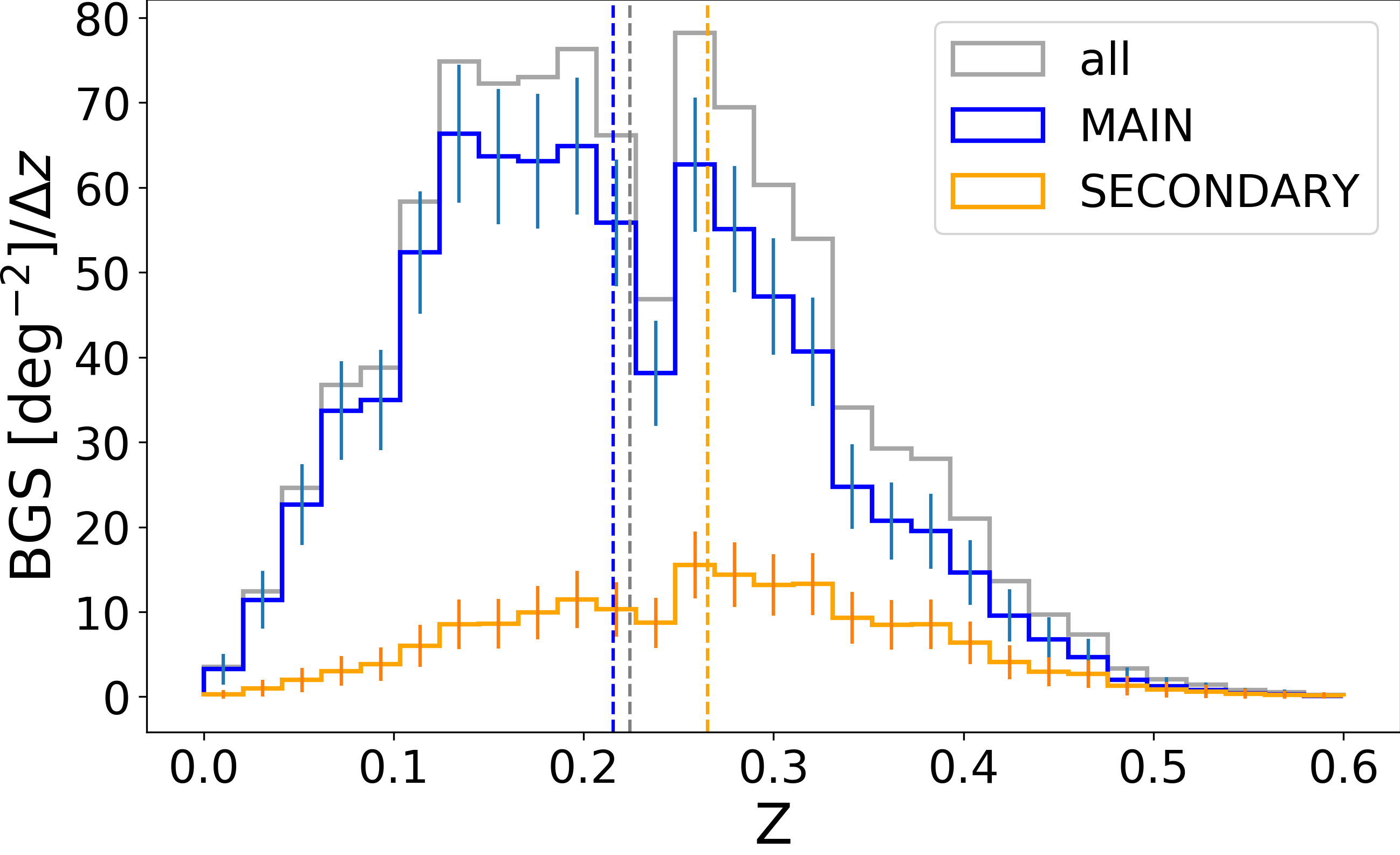}
\caption{{ \bf Left:} Target density of BGS as a function of $r$-band magnitude in DECaLS DR8. The red dashed line at $19.5$ marks the boundary between the BRIGHT and FAINT selections. { \bf Right:} Redshift distribution of DECaLS DR8 BGS targets cross-matched with GAMA DR4 in redshift bins of $\Delta z = 0.02$. The solid blue, orange and gray lines show the BRIGHT, FAINT and combined samples respectively. The dashed lines show the means of the samples at $0.21$ (BRIGHT), $0.26$ (FAINT) and $0.22$ (combined). Error bars for BRIGHT and FAINT distributions are Poisson. Because GAMA has a magnitude limit of approximately $r=19.8$ in (SDSS DR7) $r$-petrosian \citep{2009ApJS..182..543A,2012yCat..74130971D}, only $30$ per cent of the FAINT sample appears in GAMA and the mean redshift is expected to be somewhat higher. While for the BRIGHT sample, the GAMA completeness is as high as $95$ per cent.}
\label{fig:bgs_counts}
\end{center}
\end{figure}

\section{Conclusion} 

We have presented the BGS target selection from LS DR8 divided into two samples, the BRIGHT and FAINT samples.  The BRIGHT sample, which will have a higher fiber-allocation priority, comprises $\sim$800 targets/deg$^2$. The FAINT sample, which will be assigned at lower priority, comprises $\sim$600 objects/deg$^2$. Both samples undergo the spatial and photometric cuts outlined by the MASKBITS and the number of observations required, and by Equations\,\ref{eq:color_gr}-\ref{eq:fmc}  respectively, passed the {\it Gaia} based star-galaxy classification and our quality cuts. Fig.~\ref{fig:bgs_counts} shows the BGS target density as a function of $r$-band magnitude 
and the $N(z)$ of BGS targets cross-matched to GAMA DR4 \citep{2012yCat..74130971D, 2015MNRAS.452.2087L, 10.1093/mnras/stx3042}.
The preliminary selection described in this note is public\footnote{Available at \niceurl{https://data.desi.lbl.gov/public/ets/target/catalogs/} and detailed at \niceurl{https://desidatamodel.readthedocs.io}}.


\section*{Acknowledgements}
OR-M is supported by the Mexican National Council of Science and Technology (CONACyT) through grant No. 297228/440775 and funding from the European Union’s Horizon 2020 Research and Innovation Programme under the Marie Sklodowska-Curie grant agreement No 734374. This research is supported by the Director, Office of Science, Office of High Energy Physics of the U.S. Department of Energy under Contract No. DE–AC02–05CH1123, and by the National Energy Research Scientific Computing Center, a DOE Office of Science User Facility under the same   contract; additional support for DESI is provided by the U.S. National Science Foundation, Division of Astronomical Sciences under Contract No. AST-0950945 to the NSF’s National Optical-Infrared Astronomy Research Laboratory; the Science and Technologies Facilities Council of the United Kingdom; the Gordon and Betty Moore Foundation; the Heising-Simons Foundation; the French Alternative Energies and Atomic Energy Commission (CEA); the National Council of Science and Technology of Mexico; the Ministry of Economy of Spain, and by the DESI Member Institutions.  The authors are honored to be permitted to conduct astronomical research on Iolkam Du\textquotesingle ag (Kitt Peak), a mountain with particular significance to the Tohono O \textquotesingle odham Nation. 

\bibliography{biblio}

\end{document}